# Asynchronous Routing for Multipartite Entanglement in Quantum Networks

Chenliang Tian
Department of Computer Science and Engineering
Washington University
St. Louis, USA
chenliang.t@wustl.edu

Zebo Yang
Department of Electrical Engineering and Computer Science
Florida Atlantic University
Boca Raton, USA
yangz@fau.edu

Raj Jain
Department of Computer Science and Engineering
Washington University
St. Louis, USA
jain@wustl.edu

Ramana Kompella
Quantum Lab
Cisco
Santa Monica, USA
rkompell@cisco.com

Reza Nejabati
Quantum Lab
Cisco
Santa Monica, USA
rnejabat@cisco.com

Eneet Kaur
Quantum Lab
Cisco
Santa Monica, USA
ekaur@cisco.com

Aiman Erbad
Department of Computer Science and Engineering
Qatar University
Doha, Qatar
aerbad@qu.edu.qa

Mounir Hamdi
Department of Science and Engineering
Hamad Bin Khalifa University
Doha, Qatar
mhamdi@hbku.edu.qa

Mohamed Abdallah
Department of Science and Engineering
Hamad Bin Khalifa University
Doha, Qatar
moabdallah@hbku.edu.qa

*Abstract*—In quantum networks, one way to communicate is to distribute entanglements through swapping at intermediate nodes. Most existing work primarily aims to create efficient two-party end-to-end entanglement over long distances. However, some scenarios also require remote multipartite entanglement for applications such as quantum secret sharing and multi-party computation. Our previous study improved end-to-end entanglement rates using an asynchronous, tree-based routing scheme that relies solely on local knowledge of entanglement links, conserving unused entanglement and avoiding synchronous operations. This article extends this approach to multipartite entanglements, particularly the three-party Greenberger-Horne-Zeilinger (GHZ) states. It shows that our asynchronous protocol outperforms traditional synchronous methods in entanglement rates, especially as coherence times increase. This approach can also be extended to four-party and larger multipartite GHZ states, highlighting the effectiveness and adaptability of asynchronous routing for multipartite scenarios across various network topologies.

## I. Introduction

The emergence of quantum computing has heralded new advancements in communication networks [1]. Rooted in quantum mechanics, quantum networks offer enhanced security and computational capabilities beyond those of classical systems. Key applications include quantum key distribution (QKD), distributed quantum computations, and quantum teleportation [2].

A key requirement for these technologies is establishing remote entanglement over long distances. Quantum repeaters have been developed to tackle the issue of long-distance quantum entanglement degradation [3]. These repeaters utilize entanglement swapping and fusion measurements to facilitate two-party and multipartite entanglement [4] [5] [6]. Such technology extends entanglement across multiple repeaters, creating long-distance end-to-end or multi-party connections.

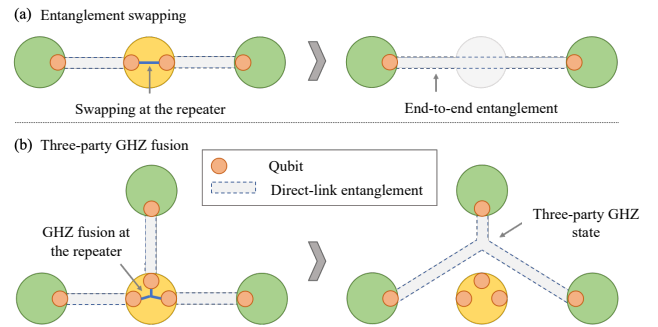

Fig. 1. Illustration of swapping and GHZ fusion.

Existing entanglement routing schemes—commonly referred to as synchronous or two-phase protocols—typically consist of two main phases: the entanglement generation phase and the entanglement distribution phase [4] [5] [6] [7]. The generation phase creates entanglement between directly connected quantum nodes through quantum operations on their qubits, forming a primary layer of entangled pairs across adjacent nodes – often called instant topology. In contrast, the underlying network of optical links is known as the physical topology. This phase establishes the foundational links or building blocks for broader network-wide entanglements. Once the instant topology is established, the distribution phase begins, aiming to extend entanglements across the network to nodes that are not directly connected. This phase involves entanglement swapping and fusion measurements, as shown in Fig. 1, with Fig. 1(b) illustrating a three-party Greenberger–Horne–Zeilinger (GHZ) state as an example. These processes allow connecting two or more entangled pairs via an

This work was supported in part by Cisco University Research Grant #98690499 and the Qatar Research, Development, and Innovation (QRDI) Academic Research Grant #ARG02-0415-240191. The statements made here are solely the responsibility of the authors.

intermediate node, effectively linking distant nodes within the network.

These two phases are typically performed in synchronized time slots under global link-state knowledge, where each repeater knows which direct-link entanglements succeeded after the generation phase. This allows centralized path selection for swapping or fusion during the distribution phase. To ensure the availability of fresh direct-link entanglement for each distribution phase, these synchronous protocols typically reset the network state at the end of every time slot and repeat the two-phase cycle in the next one. Each complete run of the two phases is considered a single timeframe. If, within a given timeframe, no connected path exists between the request nodes or if swapping or fusion attempts fail to establish end-to-end entanglement, the network advances to the next synchronous round, and the entire process is repeated. Although this global synchronization ensures, the repeated rests deplete the available entanglement within each timeframe, resulting in a lower end-to-end entanglement rate.

To address these limitations, we developed a family of asynchronous entanglement routing (AER) protocols in [8] and [9] that improve the end-to-end throughput and allow rates to scale with extended coherence times, outperforming synchronized methods. AER uses instant topology in a fully distributed manner, eliminating the need for synchronization. This instant topology generally forms a tree-like configuration, such as a destination–oriented directed acyclic graph (DODAG) [10]. Similar structures have been well researched in the context of classical lossy wireless networks, which also experience connection losses. In a quantum setting, this corresponds to entanglement loss in the instant topology, with the loss probability determined by both the physical link quality and other noise-induced decoherence effects.

While the AER scheme has proven effective for two-party end-to-end entanglement, there is a growing need to support multipartite entanglement among multiple remote endpoints. Such capability is essential for enabling applications like quantum-safe multi-party computation and conference key agreement, where a shared secret key must be established simultaneously among all participants. A practical approach is to employ multipartite entanglements, for example, Greenberger-Horne–Zeilinger (GHZ) states among the endpoints to realize secure multi-party key exchanges [11].

In this paper, we extend the DODAG-based AER framework and propose multipartite asynchronous entanglement routing (MAER) to enable the distributed generation of multipartite entanglements across distant nodes. MAER is designed to operate in a fully distributed and asynchronous manner, using only local link-state information while maintaining scalability across large quantum networks. We also demonstrate MAER's performance advantages over synchronous approaches. While Meignant et al. (2019) [6] construct multipartite entangled states under idealized, globally informed synchronization, MAER achieves superior performance without requiring global timing coordination or centralized control.

In optics-based repeaters, the success probability for an $n$-qubit fusion operation decreases exponentially, following a rate of $1/2^n$ [4]. This implies that minimizing the number of higher-order fusion gates is advantageous. Specifically, for GHZ states, the MAER protocol requires only a single n-fusion operation to establish an n-party state. This approach is tailored to GHZ-type entanglement and may not directly generalize to generating arbitrary multipartite states such as graph states or cluster states, which typically require more elaborate fusion or entangling procedures.

We assess MAER's performance through simulations across various network configurations. These simulations show consistent improvements in end-to-end entanglement rates across multiple topologies, such as grid, barbell, and random graphs. The advantage of the MAER scheme becomes particularly pronounced when quantum memories possess longer coherence times, as MAER can retain and utilize stored entanglement across multiple time slots without synchronization resets. Consequently, as memory technology advances and coherence times extend, the achievable entanglement rates under MAER continue to improve, further widening the performance gap over synchronous schemes. Moreover, a tree-like instant topology, such as a DODAG, facilitates interconnection among multiple network layers via root nodes. These results suggest that the proposed protocols can play a key role in realizing scalable quantum internetworking as quantum memory and repeater technologies mature.

Recently, an AER-based extension, the DODAG-X protocol [7], was proposed to enhance multipartite entanglement distribution across multiple receivers. This approach inherits the asynchronous DODAG structure introduced in AER and focuses on optimizing the efficiency of entanglement dissemination within a fixed routing topology. In contrast, the proposed MAER generalizes this framework to support fully asynchronous multipartite routing, where both topology formation and entanglement generation proceed concurrently without global synchronization.

The rest of the paper is organized as follows. Section II introduces the fundamental concepts that underpin MAER and multipartite entanglement routing in quantum networks. Section III presents the details of implementing the MAER protocol using DODAG. Section IV describes the simulation setup and evaluates the protocol's performance under various configurations. Section V concludes the paper.

## II. PRELIMINARIES

This section presents fundamental concepts that underpin the study of routing for multipartite entanglement in quantum networks.

### A. Direct-link Entanglement

As mentioned, quantum repeaters have been designed to expand the reach of entanglement generation. They use Bell-state measurement (BSM), i.e., entanglement swapping, to project two entangled pairs into a distant entangled state between nodes that are not directly connected. BSMs have been successfully implemented in various physical systems and are commonly used in quantum networks to link shorter entanglement chains into longer-distance clusters [12]. Quantum capabilities at network nodes during the current noisy intermediate-scale quantum (NISQ) era benefit from these probabilistic BSMs, which accommodate losses in optical fibers and the limitations of quantum hardware [1].

As quantum networks evolve beyond two-party communication, the growing demand for multipartite entanglement has motivated the use of fusion measurements, which go beyond swapping and enable the creation of GHZ states among multiple nodes. GHZ fusion measurements build on the concept of BSM by merging multiple entangled states into a single GHZ state, facilitating multipartite entanglement among nodes without direct links. This advancement necessitates new routing protocols for multi-party rather than two-party routing scenarios.

Moreover, it is important to note that a classical network is always assumed to underpin the quantum network, handling routing computations and disseminating routing information. Essentially, each node in the quantum network communicates via a classical network, aligning with the principles of local operations and classical communications (LOCC). This setup is typical in hybrid quantum-classical networks and is standard in quantum routing techniques. Once candidate paths are established via classical routing, repeaters perform entanglement swapping or fusion measurements alongside ongoing classical signaling. After each Bell-state measurement or fusion operation, the measurement outcomes are communicated through the classical network so that the end nodes can apply the appropriate Pauli corrections and complete the entanglement distribution. These message exchanges introduce classical latency proportional to network distance and hop count.

In our simulation model, classical signalling delays are not explicitly accounted for, as the focus is placed on quantum-layer timing and coherence effects. This simplification is consistent with prior studies [4] [5], which assume that classical communication time is negligible compared with the timescale of entanglement generation and swapping. Nevertheless, in practical implementations, classical feedforward for Bell-state measurement outcomes and routing coordination would introduce additional latency, which may become significant in long-haul or high-speed networks. Incorporating these classical communication effects remains an important direction for future work.

*B. Challenges in Entanglement Routing*

With repeaters enabling entanglement between remote nodes, there is a need for routing protocols that efficiently identify paths to establish entangled connections among arbitrary nodes within a network. However, quantum networks in the NISQ era present unique challenges that may not be found in classical networks, necessitating either adaptations or new designs tailored for quantum network environments:

- Entanglement as A Communication Resource: Unlike classical networks, where data packets are transmitted and relayed through intermediate nodes from the source to the destination, quantum networks utilize end-to-end or remote multipartite entanglement established via intermediate nodes. Such entanglement serves as a communication resource for transferring qubits (via teleportation) or exchanging secret keys (through QKD). This approach enhances security, as data does not physically traverse the network. However, this additional layer of entanglement distribution introduces unique challenges. For instance, while qubits can be temporarily stored in quantum memories for synchronization, they cannot be amplified or regenerated like classical signals due to the no-cloning theorem, which prohibits duplicating unknown quantum states.

- Hardware Imperfections: Quantum hardware in the NISQ era remains inherently noisy and unreliable, which significantly affects core operations in quantum networks, including direct-link entanglement generation, entanglement swapping, and fusion measurements. These operations are probabilistic and subject to failure due to factors such as gate infidelity, photon loss, mode mismatch, and limited detector efficiency. Consequently, establishing and maintaining stable entanglement links is often inefficient and unreliable. Routing protocols must be designed with these physical-layer imperfections in mind, often by modeling success probabilities for each operation and incorporating error-tolerant mechanisms or redundant paths to increase reliability in practical implementations.

- Decoherence and Photon Loss: Quantum information is inherently time sensitive due to decoherence, which causes stored quantum states to lose coherence through interactions with their environment. In quantum networks, decoherence primarily affects stationary qubits held in quantum memories. In contrast, photon loss occurs during transmission through optical fibers or free-space channels and represents a separate source of error. Consequently, networking protocols must ensure that entanglement generation, storage, and routing operations are completed within the available coherence time of the memory elements, while also accounting for transmission losses along optical links.

- Network Information Propagation: In a quantum network, routing decisions may depend on the comprehensive knowledge of the instant topology, meaning that every node must know the status of all direct-link entanglement links before routing can commence. However, as the network scales, disseminating this information across all nodes becomes increasingly time-consuming. Due to the limited coherence time of entanglement links, some links may have already decohered by the time global link-state information is fully propagated. Therefore, there is a need for routing schemes that do not rely on global knowledge of the current topology but instead use information about adjacent links to perform pathfinding or routing in a distributed manner.

- Network Scalability: As quantum networks expand to accommodate a growing number of users and increasingly complex applications, maintaining overall performance and reliability becomes more challenging. Larger networks demand more entanglement links, greater coordination among nodes, and more classical communication overhead to support entanglement routing and management. These scaling factors amplify the effects of decoherence, operation delays, and cumulative errors, especially under hardware constraints. Effective scalability requires routing protocols that efficiently manage entanglement resources across large topologies, minimize reliance on global knowledge, and dynamically adapt to changes in network size and structure while preserving the integrity of quantum information.

## C. The Routing Problem and Existing Work

As mentioned in Section I, it is common in quantum network entanglement routing schemes to structure time into discrete slots, each divided into two phases. With that, we can formulate the entanglement routing problem with a graph $G(V, E)$ that represents the physical structure of the quantum network. In this graph, each node $v \in V$ acts as a repeater, and each edge $e \in E$ indicates a physical channel connecting two adjacent repeaters. The instant topology, $G'(V', E')$ derived from $G(V, E)$, where $E'$ represents the direct-link entanglement links and $V'$ includes nodes interconnected by these links. Each edge $e$ supports entanglement generation through a qubit pair between neighboring nodes. Each node possesses a limited number of qubits. For simplicity, only a single qubit on each side of a physical link is assumed, as shown in Fig. *1*.

Due to operational uncertainties, entanglement generation on a direct link $e$ has a success probability denoted as $p(e)$, which depends on factors such as physical distance and channel transmissivity, and photon loss. Similarly, the success probability of entanglement swapping at a repeater node, denoted by $q(v)$, is primarily determined by the performance of the BSM apparatus. Key influences include detector efficiency, photon indistinguishability, memory retrieval fidelity, and channel loss. In practice, linear-optical BSMs are inherently probabilistic and can succeed with a probability up to 50% even under ideal conditions, with this probability further reduced by imperfections such as photon loss, finite coherence time, and synchronization errors.

For analytical tractability, and consistent with prior works on repeater-based entanglement routing [4] [5] [8], we assume uniform probabilities $p$ and $q$ for entanglement generation and swapping across the network, respectively. This simplification is reasonable for networks composed of homogeneous optimal links and repeater hardware. More heterogeneous configurations could instead assign $p(e)$ and $q(v)$ based on link- and node-specific physical parameters to capture variations in hardware performance and physical-layer conditions in future studies.

Each entanglement is also assumed to have a constant coherence time, $T_{CO}$, reflecting how long it can remain stable without significant degradation. Assuming a single connection request at a time, the end-to-end entanglement rate, denoted by $\xi$, quantifies the number of remote entanglements formed per unit time T, which must not exceed $T_{CO}$. The time for both phases of operation is counted as a single unit, set to $T = T_{CO}/m$, where m is an integer and $m \geq 1$. For example, if $T_{CO} = m$, it suggests that the coherence time spans n unit times in that simulation.

As discussed earlier, finding a path with the global knowledge approach is well studied, but often impractical due to the extended time required to disseminate link-state information across the network. Most existing studies, Most existing studies, such as those by Shi and Qian (2020, Q-CAST) [5], Meignant et al. (2019) [6], and Negrin et al. (2024) [7], adopt this paradigm, rely on such global knowledge, employing synchronization mechanisms to propagate the instant topology after the first phase. This allows straightforward path identification using shortest-path algorithms during the second phase. In contrast, relying on local knowledge of the instant topology is more feasible in practice, though it complicates pathfinding due to limited information about the overall network topology [8]. Nonetheless, once a path consisting of $k$ edges and $k - 1$ repeaters is successfully established after the completion of several time slots, and a remote entanglement between the end nodes can be achieved with a probability of $q^{(k-1)}$.

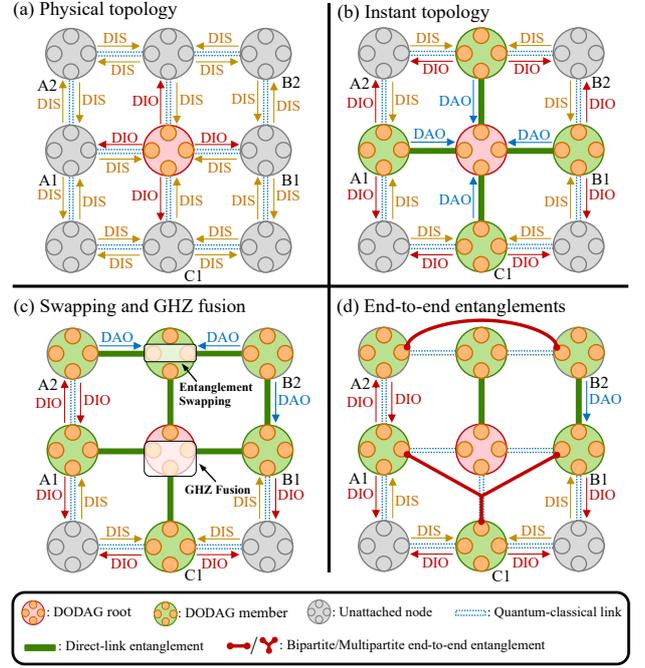

Fig. 2. DODAG-based entanglement routing and its update.

## III. MULTIPARTITE ASYNCHRONOUS ENTANGLEMENT ROUTING

This section introduces MAER.

### A. Tree-based Asynchronous Routing

The MAER scheme builds upon the framework introduced in our prior work [8] [9]. In this scheme, nodes manage and update the instant topology in a distributed manner by exchanging DODAG messages over classical channels, allowing the instant topology to evolve dynamically as a tree. Analyses involving a spanning tree in this context are also discussed in [8].

A DODAG in MAER is structured with all edges directed towards a designated root node. The root node can be pre-assigned based on network role (e.g., a data center or service node) or dynamically elected based on network metrics such as degree centrality or memory availability. In our simulations, the geometric center of the network topology is designated as the root to minimize the average hop distance to all nodes.

The DODAG construction rule ensures loop freedom. Each node selects only a parent with a strictly lower rank value. Because ranks monotonically decrease toward the root, cyclic dependencies cannot occur. This mechanism is analogous to the classical Routing Protocol for Low-Power and Lossy Networks (RPL), where ordered ranks prevent loops.

The DODAG is constructed through the exchange of control messages, and in the quantum setting, nodes use rank values to represent their logical distance from the root. These ranks facilitate the selection of optimal routes for entanglement distribution and simplify path management by enabling distributed routing decisions centered around the root node. Each node, including consumers, maintains only local knowledge of its neighboring links and parent rank values within the DODAG. Nodes neither require nor store global network topology information.

Fig. *2* demonstrates the growth of a DODAG instance from a designated root node under the MAER protocol, which establishes end-to-end entanglement between end nodes, such as (A1, B1) and (A2, B2, C2). Nodes already included in the DODAG are depicted in green, while nodes not included are shown in grey. Nodes within the DODAG can serve as parents to other nodes during topology expansion.

As shown in Fig. *2*(a), unjoined (gray) nodes initiate the joining process by sending DIS (DODAG Information Solicitation) messages to their neighboring nodes. Upon receiving a DIO (DODAG Information Object) message from a neighbor already in the DODAG, the grey node replies with a DAO (Destination Advertisement Object) message to formally request inclusion. The neighbor then assigns a rank to the requesting node and integrates it into the DODAG. Only nodes that are not yet part of the DODAG broadcast DIS messages. Once a node successfully joins and obtains a rank, it stops transmitting DIS. Instead, it responds to new solicitations with DIO messages, allowing the tree to expand outward from the root.

If no DIO messages are received, the unjoined node continues broadcasting DIS messages until a valid response is received or another neighboring node admits it. As the DODAG evolves (Fig. *2*(b)-(d)), routing paths gradually form by selecting parent nodes with the lowest rank.

In the baseline MAER configuration, the DODAG expands outward from a designated root node, which gradually attracts join requests from consumers and intermediate nodes. Alternatively, one could imagine a consumer-initiated inward expansion, in which each end node independently attempts to connect to the root. In that case, multiple partial trees would form concurrently and merge when they encounter a common ancestor. While this approach could reduce the initial setup latency for sparsely connected nodes, it introduces coordination overhead to resolve rank conflicts and maintain acyclicity during tree merging. The present MAER formulation avoids this complexity by allowing all joining nodes, including request nodes, to send solicitation (DIS) messages only toward existing DODAG members, ensuring consistent rank ordering from the root outward.

A further extension involves multiple root nodes (multi-root MAER), which could represent distinct service centers or access points. In this configuration, independent DODAGs would form around each root and could later be fused through inter-root entanglement or GHZ-fusion operations to link otherwise separate clusters. While multi-root designs may improve robustness and load balancing, they also complicate coordination and rank assignment across overlapping regions. Investigating such multi-root synchronization and inter-tree fusion mechanisms is an interesting direction for future work.

*B. Topology Update for Multipartite States*

The root node does not require prior global knowledge of consumer locations. Instead, end nodes periodically broadcast local join (DIS) messages over classical channels, which are relayed until received by a DODAG member. Through subsequent DIO and DAO exchanges, the root node gradually learns which nodes have joined and their relative ranks, enabling the formation of a dynamic tree without centralized location tracking. Classical communication for these control messages occurs asynchronously alongside entanglement generation attempts.

Consider Alice, Bob, and Charlie, located at $A_1$, $B_1$, and C1, respectively, in Fig. *3*, who aim to establish a three-party GHZ state but lack direct optical links connecting them. As the DODAG expands from the root, these end nodes attempt to join the tree as described above. Each node continuously transmits DIS messages until successfully admitted into the DODAG, as illustrated in Fig. 3(a) and (b).

After joining the DODAG, each of the three end nodes establishes an end-to-end entanglement with the root node through a sequence of entanglement-swapping operations, if required. Once these operations succeed, three distinct entanglement links are formed between the root node and the end nodes, as indicated in Fig. *3*(b). The root node then performs a fusion measurement on its three qubits. If the fusion succeeds, a remote GHZ state is generated among Alice, Bob, and Charlie, as shown in Fig. *3*(d). This GHZ state can then be utilized for various quantum applications, such as multi-party QKD or distributed computation.

After the requesting end nodes consume a GHZ entanglement, the network updates accordingly to accommodate subsequent connection requests. Only the entanglement links involved in the established GHZ state are consumed, while unused direct-link entanglement remains available for later time slots or until lost due to decoherence. Notably, the proposed scheme supports both two-party and multipartite entanglement requests and requires only a single n-fusion operation to establish an n-party GHZ state.

*C. Path Selection*

Path selection becomes relatively flexible once the end nodes are integrated into the DODAG. Each node selects a parent with a lower rank to facilitate entanglement swapping toward the root node. If there is only one available parent, the node simply chooses it. When multiple parents are available, the node prefers the one with the lowest rank, as this corresponds to the shortest logical distance to the root. In cases where multiple parents share the same minimum rank, the node can arbitrarily choose among them, as we assume no difference in link quality or reliability.

It might initially appear necessary for every pair of end nodes to establish independent paths to the root, since only one direct-link entanglement can exist between any two nodes. Fig. *3* illustrates that this approach can be wasteful. For instance, $B_2$ could, in principle, create an end-to-end link with the root through $A_1$, while $A_2$ does so via node $L$. However, such routing would consume the direct-link entanglement between $A_1$ and the root, which prevents the successful completion of the three-party request among $A_1$, $B_1$, and $C_1$, and would additionally deplete the root's direct link to the node $L$. A more efficient alternative is to allow local aggregation, as shown in Fig. 3(c). Node $L$ is already the parent of $A_2$ and $B_2$. Then, node $A_2$ and $B_2$ can establish their end-to-end entanglement directly through their common parent node $L$, without independently connecting to the root. This local-path strategy preserves valuable direct-link entanglement resources, enabling simultaneous fulfillment of other requests such as the three-party state $(A_1, B_1, C_1)$.

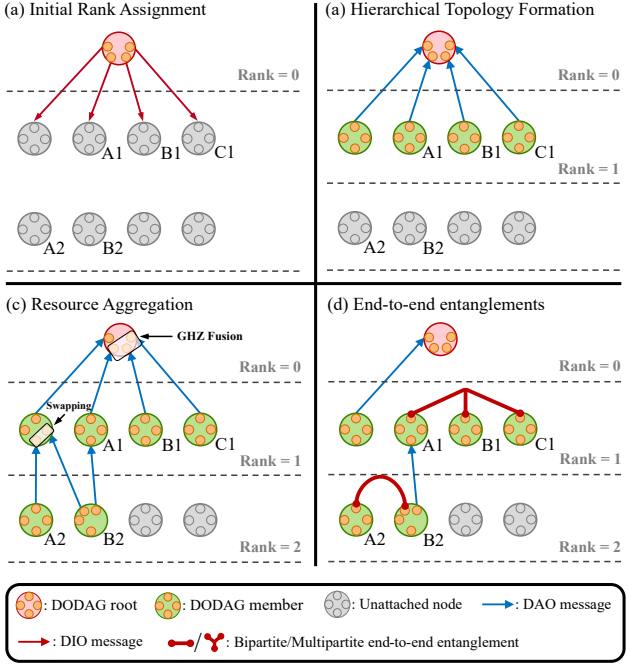

Fig. 3. DODAG-based bipartite and multipartite entanglement routing.

Fig. *4* illustrates two distinct approaches for generating three-party GHZ states. The first adopts the conventional method, in which a common parent performs a GHZ-fusion measurement to entangle $(A_1, B_1, C_1)$, as shown in Fig. *4*(c). The second employs a Hadamard-plus-CNOT fan-out principle [13] to generate the GHZ state $(A_2, B_2, C_2)$ without centralized fusion. As shown in Fig. *4*(c), the root node first performs an entanglement-swapping operation to connect the Bell pairs $(A_2, \text{Root})$ and $(\text{Root}, B_2)$, thereby creating an end-to-end Bell pair between $A_2$ and $B_2$. The parent node $A_2$ then applies a Hadamard gate to its qubit entangled with $C_2$, generating the superposition $(|0\rangle + |1\rangle)/\sqrt{2}$. Next, $A_2$ performs a CNOT operation using this qubit as the control and its qubit entangled with $B_2$ as the target. This sequence locally extends the two Bell pairs into a three-party GHZ state, including $A_2$, $B_2$, and $C_2$ without requiring additional swapping.

In our MAER scheme, both centralized and distributed strategies are employed for multipartite entanglement generation, depending on network conditions and resource availability. When multiple nodes share a common parent, the parent can act as a local fusion center, performing the necessary entanglement-swapping or Hadamard-plus-CNOT fan-out operations to create GHZ states without involving the root. Alternatively, when the participating nodes are located in different DODAG branches, the root node performs GHZ fusion to establish the multipartite entanglement across distant clusters.

## IV. EVALUATION

This section assesses the proposed scheme by detailing the simulation settings and presenting the results. To contextualize the evaluation, recall that in synchronous routing schemes, such as [5] and [6], all nodes operate in globally synchronized time slots. In contrast, the proposed MAER executes entanglement generation and swapping independently across nodes without global timing coordination.

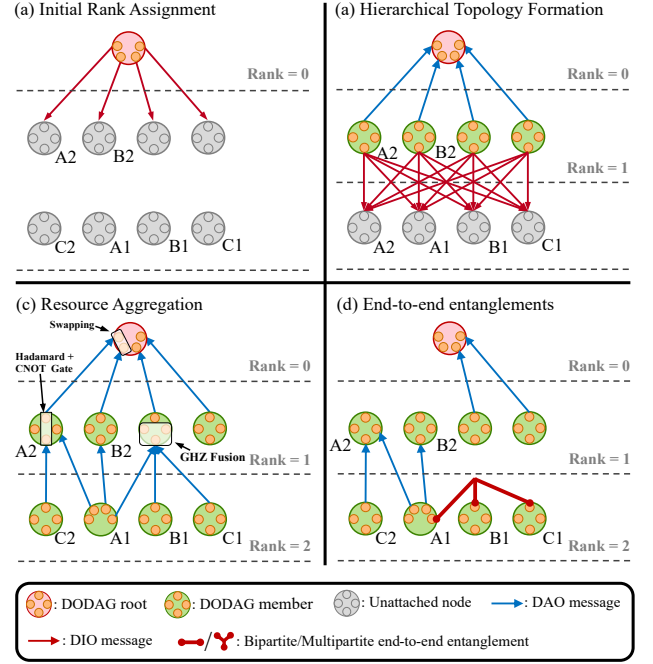

Fig. 4. Two approaches for generating three-party GHZ states.

### A. Simulation Setting

In the simulation, we assume fixed success probabilities for direct-link entanglement generation and repeater operations. Specifically, each direct link succeeds with probability $p$, and each repeater operation (e.g., entanglement swapping, GHZ fusion, or Hadamard-plus-CNOT gate) succeeds with probability $q$. To align with prior synchronous approaches for comparison, we define a unit of time as a time slot. The coherence time $T_{co}$ (often denoted $T_2$) represents the duration over which stored entanglement remains stable [14].

We further assume that attempts to establish direct-link entanglement and to operate a repeater are independent. Once a path of length $k$ is established within the instant topology, the probability of successfully generating a GHZ state along that path is given by $\xi = q^{(k-1)}$. Because an entanglement attempt is performed per unit time, this success probability directly corresponds to the expected end-to-end entanglement rate per unit time.

We evaluate MAER with three consumers (end nodes) per experiment. Unless noted otherwise, we use the graph-theoretic shortest-path hop distance rather than the Euclidean layout distance. For a consumer triple $\{u, v, w\}$, we define the triplet distance d = $\frac{1}{3}\bigl(D(u,v) + D(v,w) + D(w,u)\bigr)$, where $D(\cdot,\cdot)$ is the shortest-path hop count on the physical topology. When plotting performance versus distance $d$, we target a desired value $d^\star$ and sampling consumer triples uniformly at random under the constraint $d \in [d^\star - \Delta, d^\star + \Delta]$, with tolerance $\Delta = 1$ hop (unless otherwise specified).

For each parameter combination and topology, we draw. $N = 100$ independent consumer triples and reports the mean end-to-end entanglement. Since centrally located nodes benefit from higher path diversity (larger network min-cut) than peripheral ones, we randomly sample at a fixed $d^\star$ averages over these effects and prevents bias toward particularly favorable geometries.

Each simulation runs for 20,000 iterations to compute the average multipartite entanglement rate. Different physical topologies are generated by NextworkX [15], including a grid network, a barbell network with a single backbone link, and a random graph in which each node has an average of 4 links, each consisting of 100 network nodes. At the beginning of each simulation, the central node of the physical topology is designated as the root, and a three-user request is then initiated by randomly selecting nodes located at the specified graph distance. Each selected node attempts to join the DODAG by establishing a direct link with existing members. As successful links form, the DODAG dynamically expands, and rank values are assigned to newly joined nodes based on their distance from the root.

As the DODAG evolves, it attempts to form a connected structure that includes all three end users. If no such connection is established within a unit of time, the GHZ entanglement generation rate for that period is recorded as zero. Suppose at least one valid path is identified. In that case, the shortest path, potentially through a common parent node rather than the root, is selected and used to compute the corresponding success probability for multipartite entanglement generation.
In our simulations, we abstract away entanglement fidelity, assuming that all successful direct-link entanglement attempts yield usable pairs of uniform quality. While fidelity maintenance is crucial for practical implementations, it is beyond the scope of this study and is left for future study.

*B. Results*

Unless otherwise specified in the caption, Fig. *5*(a)-(b), and Fig. *6*(a) use the grid topology (10×10, 100 nodes) and consumer triples sampled by the procedure above with $\Delta = 1$ and $N = 100$ trials per point.

As shown in Fig. *5*, the performance of traditional synchronous protocols [5] [6] is labelled as "Syn" in the charts. The result shows that the proposed MAER protocol based on DODAG achieves higher end-to-end entanglement rates for remote three-qubit GHZ generation than synchronous methods, particularly when the coherence time exceeds 1. The purple-dashed lines, which emerge as the coherence time approaches infinity, represent the estimated upper bounds on the entanglement rates achievable by MAER under the simulation settings. The end-to-end entanglement rate achieved by MAER increases steadily with longer coherence times. This suggests that as quantum technology advances, enabling quantum memory to maintain entangled states for extended periods, the advantage of asynchronous routing will continue to grow relative to synchronous schemes.

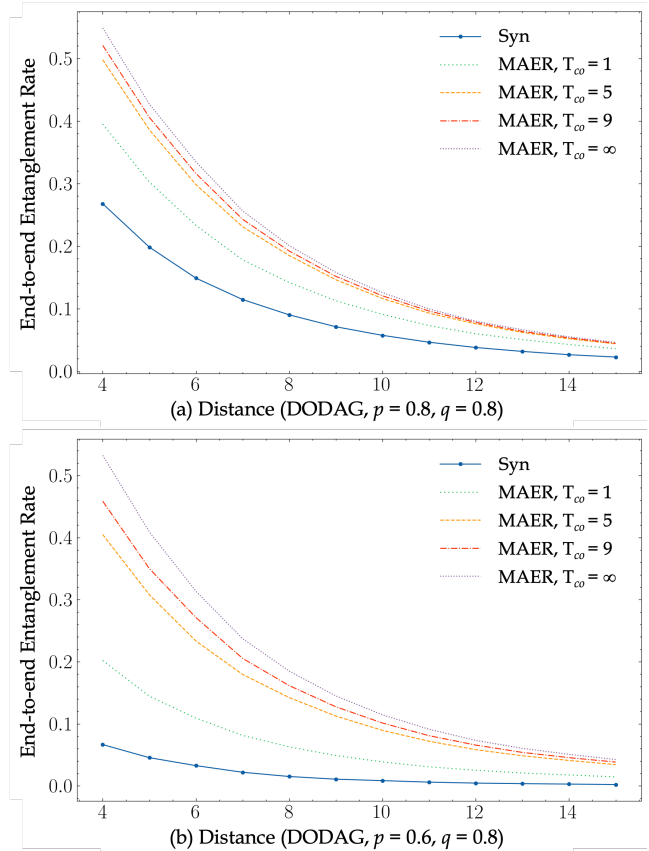

Fig. 5. Rate vs. graph distance with varying coherence times.

To further evaluate performance under different parameter settings, we tested various combinations of the direct-link success probability $p$ and the repeater operation success probability $q$. Except for cases where the coherence time equals one and $p$ is small, MAER consistently outperforms synchronous methods across all tested combinations, as shown in Fig. *6*(a). It also demonstrates that MAER achieves a consistently higher upper bound for multipartite end-to-end entanglement rates than synchronous protocols under the same conditions. Moreover, as shown in Fig. *6*(b), the proposed MAER protocol maintains its advantage in achieving higher multipartite end-to-end entanglement rates across all tested network topologies, including a grid, a barbell, and a random graph. Each network topology used in the simulation contains 100 nodes. The grid topology is a ten-by-ten square lattice, where each interior node is connected to four neighbors. The barbell topology consists of two fully connected clusters (cliques), each with 50 nodes, connected by a single backbone link between one node in each cluster. The random graph topology is generated using the Erdős–Rényi model with 100 nodes and a connection probability of 0.04. Results for the grid and random graph are similar, likely due to their comparable average degree of approximately four under the given configuration, despite differences in their structural properties.

Considering a generalized scenario with an arbitrary number n of end nodes forming an $n$-qubit GHZ state, we conducted additional simulations to evaluate MAER's performance under varying numbers of participants. Specifically, we examined cases where the number of end nodes $n$ ranged from 3 to 7, as shown in Fig. 7. To maintain comparable spatial separation between nodes as $n$ increases, the average pairwise graph distance $d$ among end nodes was set proportionally to $n$, defined as $d = 3 \times n$. This proportional scaling prevents newly added participants from clustering near existing participants and ensures that network load and path diversity scale roughly linearly with the multipartite group size. If $d$ were held constant while increasing $n$, additional users would be placed within a fixed spatial region, leading to overlapping paths, shared repeaters, and artificial correlation between entanglement-generation attempts. While such a setting might be interesting for analysing local congestion effects, our goal here is to study MAER's scalability with system size under comparable topological spreading. Across all these simulations, the MAER protocol consistently achieved higher end-to-end entanglement rates for $n$-qubit GHZ states compared with existing synchronous approaches, for both grid (Fig. 7(a)) and random (Fig. 7(b)) topologies.

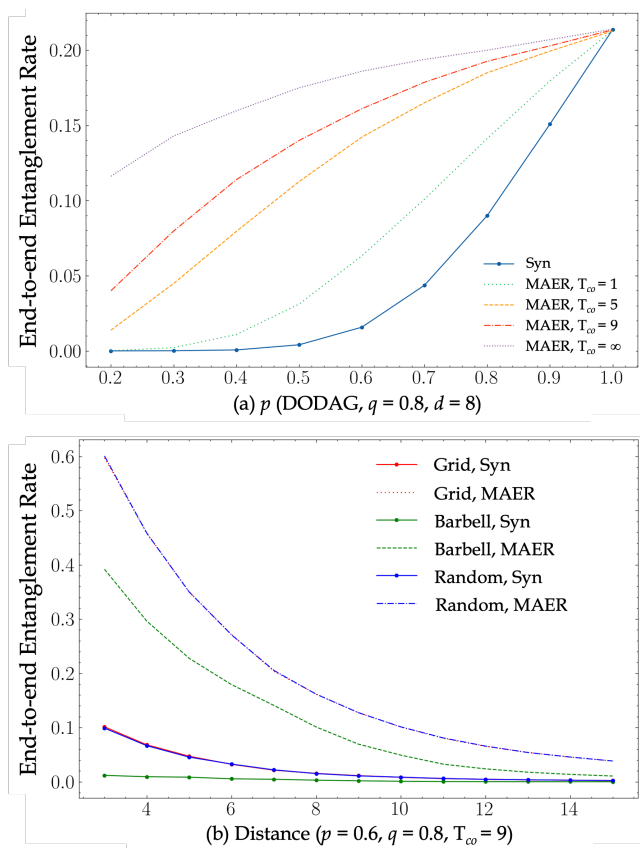

Fig. 6. Rate vs. p across various coherence times and Rate vs. d across various topologies.

Based on the simulation results and subsequent analyses, we substantiate the advantages of employing the MAER protocol with a DODAG structure for multipartite entanglement distribution. These advantages are evident across several key dimensions: (1) The end-to-end entanglement rate for remote GHZ states under MAER shows a positive correlation with increased coherence times, indicating improved performance as quantum memories become more stable; (2) MAER is adaptable to diverse network topologies—including grid, barbell, and random graphs—while consistently outperforming synchronous methods; (3) In simulations involving dynamic $n$-party GHZ generation, MAER maintains high end-to-end entanglement rates even as the number of end nodes increases, demonstrating its effectiveness in handling multipartite entanglement requests of varying sizes; (4) These properties collectively underscore MAER's strong scalability, making it well-suited for future quantum networks expected to support larger user bases and more complex quantum applications.

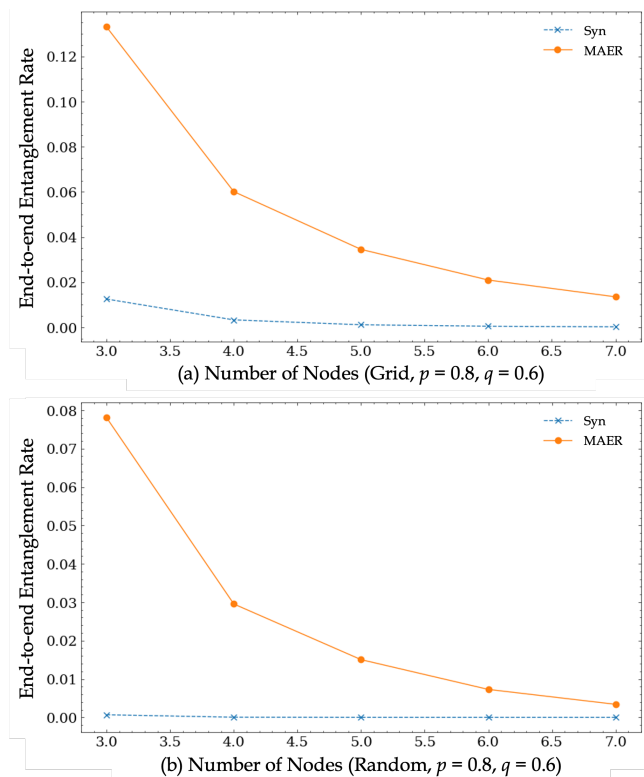

Fig. 7. Rate vs. the number of nodes and across various topologies.

## V. CONCLUSION

This study advances quantum network communications by extending asynchronous, tree-based routing schemes to support both two-party end-to-end entanglement and more complex multipartite entanglement, such as GHZ states. The proposed MAER protocol leverages only local knowledge of entanglement links, eliminates the need for synchronized operations, and conserves unused entanglement resources. Simulation results demonstrate that MAER significantly outperforms existing synchronous methods, particularly as coherence time increases. These findings highlight the adaptability and effectiveness of asynchronous routing in supporting multipartite entanglement across diverse network topologies. This enhances the practicality of quantum networks for advanced applications, such as quantum secret sharing and multi-party computation, and opens promising directions for future research.

Among these, a particularly important challenge is managing multiple simultaneous entanglement requests, which requires careful link capacity allocation and resource coordination. This aspect is not addressed in the current work and represents a valuable area for further investigation.